\documentclass[aps,twocolumn,showpacs,preprintnumbers,floats]{revtex4}

\usepackage{graphicx}% Include figure files
\usepackage{dcolumn}% Align table columns on decimal point
\usepackage{bm}% bold math
\usepackage{epsfig}

\begin{document}
\title{New solutions for the color-flavor locked strangelets}
\author{G.\ X.\ Peng$^{1,2,3}$, X.\ J.\ Wen$^2$, Y.\ D.\ Chen$^{2}$}
\affiliation{
 $^1$China Center of Advanced Science and Technology
   (World Lab.), Beijing 100080, China\\
 $^2$Institute of High Energy Physics,
   Chinese Academy of Sciences, Beijing 100039, China \\
 $^3$Center for Theoretical Physics MIT,
% Laboratory for Nuclear Science and Department of Physics, \\
% Massachusetts Institute of Technology,
      77 Mass.\ Ave., Cambridge, MA 02139-4307, USA
            }
%\date{\today}

\begin{abstract}
Recent publications rule out the negatively charged beta
equilibrium strangelets in ordinary phase, and the color-flavor locked
(CFL) strangelets are reported to be also positively charged. This letter
presents new solutions to the system equations where CFL strangelets are
slightly negatively charged. If the ratio of the square-root bag constant
to the gap parameter is smaller than 170 MeV, the CFL strangelets are
more stable than iron and the normal unpaired strangelets. For the same
parameters, however, the positively charged CFL strangelets are more stable.
\end{abstract}

\maketitle

After the acceptance of quantum chromodynamics as the
fundamental theory of strong interactions, it became extremely
significant whether a deconfined phase of matter consisting merely
of quarks would be possible. Theoretical investigations show that
strange quark matter (SQM), which is composed of $u$, $d$, and $s$
quarks, might be absolutely stable
 \cite{Witten84,Farhi,Peng00PRC61}.
% \cite{Witten84,Farhi,Peng00PRC61,wxj05prc72}.
Because small lumps of SQM, the so called strangelets, could be
produced in modern relativistic heavy-ion collision experiments,
their charge property has attracted a lot of interest
\cite{Dar}.
%\cite{Dar,JaffeRMP72,Glashow,Kent}.

Originally, SQM is believed to show up with some small positive
charge \cite{Farhi}.
In June 1997, however, Schaffner-Bielich {\sl et al.}\ demonstrated that
strangelets are most likely heavily negatively charged \cite{Schaf}.
In June 1999, it was shown that negative charge can lower the critical density
of  SQM \cite{PengPRC59}.
In July 1999, Wilczek mentioned an ``ice-9''-type transition \cite{Wilczek},
which was picked up by a British newspaper.
Not long ago, in response to public concern, an expert committee published
a report \cite{JaffeRMP72}, which got positive comments \cite{Glashow},
as well as criticisms \cite{Kent}.
In fact, the strangelets in Ref.\ \cite{Schaf} are not in $\beta$\
equilibrium which drives the system to flavor equilibrium, and
negatively charged strangelets in normal phase have been ruled out
by a recent publication \cite{Madsen00PRL85}.

Much progress has been achieved recently by the
introduction of color supperconductivity \cite{Rajagopal2001,Huangm03PRD67}.
It has been shown that bulk SQM with color-flavor locking is electrically
neutral \cite{Rajagopal01PRL86}. Immediately, Madsen found a solution
to the corresponding system equations of strangelets, where
color-flavor locked strangelets are positively charged \cite{Madsen01PRL87},
and they might be a candidate for cosmic rays beyond the GZK cutoff
\cite{Madsen03PRL90}.

 \begin{figure}[htb]
\epsfxsize=8.2cm
\epsfbox{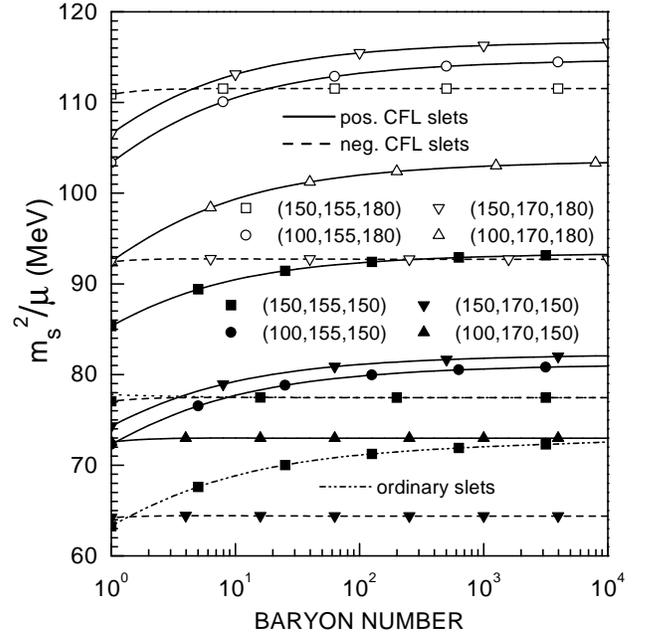}
 \caption{
The ratio of squared strange quark mass to chemical potential
for different strangelets with various parameters. The solid lines are
for the CFL strangelets in Ref.\ \cite{Madsen01PRL87}.
The dashed lines give the new solutions of CFL strangelets reported in
this letter. Unlike the previous strangelts, which are positively charged,
these new strangelets are slightly negatively charged, or nearly neutral.
Parameters are indicated as ($\Delta,B^{1/4},m_s$) in MeV.
         }
\label{figms}
\end{figure}

Very recently, it is shown that CFL phase can exist only when the
ratio of the squared strange quark mass to chemical potential,
i.e., $m_s^2/\mu$, is smaller than a critical value about
2 times the gap parameter $\Delta$ \cite{Alford04PRL92}.
It is therefore of interest to study if there is some similar
criterion for CFL strangelets given that surface effects become
quite important.
Fig.\ \ref{figms} explicitly shows the ratio
for various parameters.
% for the positively
%charged CFL strangelets in the Fig.\ 2 of Ref.\ \cite{Madsen01PRL87}.
%It is very obvious that all such CFL strangelets are within
%the paremeter range where the bulk CFL phase can exist.
The solid lines are for the CFL strangelets reported in
Ref.\ \cite{Madsen01PRL87}. These strangelets are positively charged.
At the same time, there are new solutions (the dashed lines)
which are slightly negatively charged or nearly charge-neutral,
to be discussed in detail below.
It is found that the stability of the CFL strangelets can
be judged by the ratio $\sqrt{B}/\Delta$, i.e., the square-root
bag constant to the gap parameter. If the ratio is less than
about 170 MeV, these CFL strangelets are more stable than $^{56}$Fe,
i.e., the energy per baryon is less than 930 MeV.

As done in Ref.\ \cite{Madsen01PRL87}, the thermodynamic
potential density is written as
 $\Omega=\Omega_{\mathrm{f}}+\Omega_{\mathrm{pair}}+B$.
%% Here $B$ is the normal bag contantant to keep pressure balance.
The paring contribution is
$\Omega_{\mathrm{pair}}=-3\Delta^2\bar{\mu}^2/\pi^2$\
with $\Delta$\ being the paring gap and
$\bar{\mu}=(\mu_u+\mu_d+\mu_s)/3$ being the average chemical potential
of quarks.  The normal quark contribution is
\begin{equation}
\Omega_{\mathrm{f}}
 = \sum\limits_{i=u,d,s}
     \int_0^{\nu}
     \left(\sqrt{p^2+m_i^2}-\mu_i\right)
     n^{\prime}(p,m_i,R)
      dp
\end{equation}
with the density of state $n^{\prime}(p,m_i,R)$ given
in the multi-expansion approach \cite{Balian70} by
\begin{equation} \label{tmd}
n^{\prime}(p,m_i,R)
  =g\left[
      \frac{p^2}{2\pi^2}
      +\frac{3}{R} f_{\text{S}}\left(\frac{m_i}{p}\right) p
      +\frac{6}{R^2} f_{\text{C}}\left(\frac{m_i}{p}\right)
    \right],
\end{equation}
where g=6 is the degeneracy factor for quarks, and
 the functions $f_{\text{S}}$ \cite{Berger87} and
$f_{\text{C}}$ \cite{Madsen94} are given, respectively, by
$f_{\mathrm{S}}(x)=-\mbox{arctan}(x)/(4\pi^2)$ and
$f_{\mathrm{C}}(x)=[1-(3/2)\mbox{arctan}(x)/x]/(12\pi^2)$.

%The parameter $\nu$\ is a fictional `Fermi momentum'. It is
%not the same as in the unpaired case. The CFL state has no
%Fermi momentum and the parameter $\nu$ does not fully specify
%the quark number densities. It is for convenience of
%constructing the thermodynamic potential density $\Omega$.

The common Fermi momentum $\nu$\ is a fictional intermediate
parameter. It does not fully specify the quark number density,
as it does in the unpaired case. The basic requirement
is that it can not be negative.
As a general practice, it is determined by minimizing
$\Omega$\ at fixed radius $R$, i.e., % \cite{Madsen01PRL87},
\begin{equation} \label{eqnu}
\frac{\partial\Omega}{\partial\nu}
=\sum\limits_{i=u,d,s} n^{\prime}(\nu,m_i,R)
                       \left[\sqrt{\nu^2+m_i^2}-\mu_i
                       \right]=0.
\end{equation}
The number densities $n_i\ (i=u,d,s)$ for quarks are
\begin{equation}
n_i=-\left.\frac{d\Omega}{d\mu_i}\right|_{R}
   =-\frac{\partial\Omega}{\partial\mu_i}
    -\frac{\partial\Omega}{\partial\nu}
     \frac{\partial\nu}{\partial\mu_i} \label{nk1}.
\end{equation}
Because of Eq.\ (\ref{eqnu}), the second term vanishes, while
the first term gives
$%\begin{equation}
n_i
=g\nu^3/(6\pi^2)
 +3n_{\mathrm{i,S}}/R
 +6n_{\mathrm{i,C}}/R^2
 +2\Delta^2\bar{\mu}/\pi^2,\
$%\end{equation}
where
\begin{eqnarray}
 n_{\text{i,S}}
&=& \frac{gm_i^2}{8\pi^2}
           \left[
  \phi_i-\tan\phi_i-\left(\frac{\pi}{2}-\phi_i\right)\tan^2\phi_i
           \right], \\
n_{\text{i,C}}
&=& \frac{gm_i}{16\pi^2}
           \left[
  \phi_i+\frac{1}{3}\tan\phi_i
 -\left(\frac{\pi}{2}-\phi_i\right)\tan^2\phi_i
           \right]
\end{eqnarray}
with $\phi_i\equiv\mbox{arctan}(\nu/m_i)$.

The chemical potentials $\mu_i$ and the radius $R$
are the independent state variables. For a given baryon number $A$,
one should give these quantity to fix a strangelet.
To have chemical/weak equilibrium, the chemical potentials $\mu_i$
satisfy $\mu_d=\mu_s=\mu_u+\mu_e$, maintained by reactions
such as $u+d\leftrightarrow s+u$, $u+e^-\leftrightarrow d+\nu_e$.
Because the strangelet radius is much smaller than the
Comptom wave length of electrons, the electron's number density,
and accordingly the chemical potential $\mu_e$, must be zero.
Therefore, strangelets in perfect $\beta$\ equilibrium always have
$\bar{\mu}=\mu_u=\mu_d=\mu_s\equiv\mu$.
Consequently, Eq.\ (\ref{eqnu}) gives
\begin{equation} \label{mubexp}
\mu
=\sum_in^{\prime}(\nu,m_i,R)\left.\sqrt{\nu^2+m_i^2}\right/
 \sum_in^{\prime}(\nu,m_i,R).
\end{equation}

When $R\rightarrow\infty$ and $m_u=m_d=0$,
this equation gives
$\mu=(2\nu+\sqrt{\nu^2+m_s^2})/3$
or
$\nu=2\mu-\sqrt{\mu^2+m_s^2/3}$,
which is the same as in Refs.\ \cite{Rajagopal01PRL86,Lugones2002PRD66} for
bulk CFL quark matter.
For a given baryon number $A$, one naturally has
\begin{equation} \label{eqnb}
n_{\mathrm{b}}\equiv
 \frac{1}{3}\sum_{i=u,d,s}n_i
=\frac{3A}{4\pi R^3}.
\end{equation}
To maintain mechanical equilibrium, one must require
that the pressure is zero, i.e.,
\begin{equation} \label{eqP}
P=-\Omega-\frac{R}{3}\frac{\partial\Omega}{\partial R}=0.
\end{equation}
Please note, there is an extra term when it is compared to
the normal case $P=-\Omega$. This is because of the direct
radius (or volume) dependence of the thermodynamic potential
density.

 \begin{figure}[ht]
\epsfxsize=8.2cm
\epsfbox{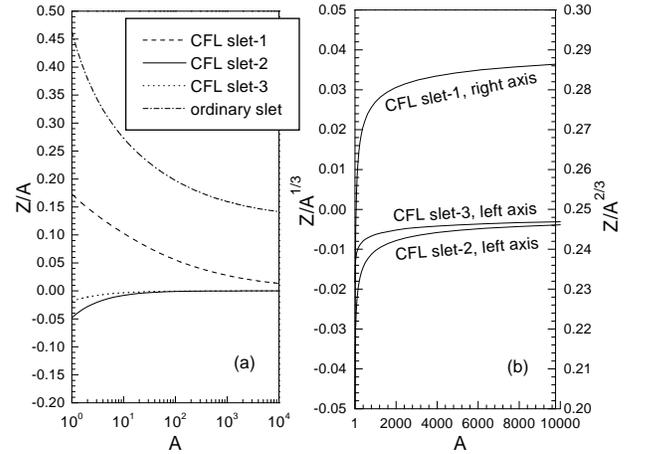}
 \caption{
Charge of strangelets. The horizontal axis is the baryon number $A$.
The vertical axis is the electric charge $Z$ to $A^{2/3}$.
There are three kinds of CFL strangelets marked with
CFL slet-1 (dashed line), CFL slet-2 (solid line),
and CFL slet-3 (dotted line). They are, respectively, charge-positive,
negative, and nearly neutral. The ordinary strangelets are also
 plotted (dot-dashed line).}
 \label{figza}
 \end{figure}

For a given baryon number $A$, one can solve the three
equations (\ref{mubexp}), (\ref{eqnb}), and  (\ref{eqP})
for $\mu$, $\nu$, and $R$. Then the overall electric charge
is $Z=V(2n_u/3-n_d/3-n_s/3)$ with $V=4\pi R^3/3$ being the volume.
%The energy density is calculated by the normal relation
%$E=\Omega+\sum_i\mu_i n_i$.
Numerical results are given in Fig.\ \ref{figza}
for parameters $\Delta=150$ MeV, $B=$ (155 MeV)$^4$, and $m_s=150$ MeV.
It is found that there are three solutions for each given baryon
number $A$. The strangelet corresponding to the first solution
(dashed line) is positively charged. It is just the one
that has been previously found in Ref.\ \cite{Madsen01PRL87}.
The strangelet corresponding to the second solution is negatively
charged (solid line), and the third solution is nearly neutral
(dotted line).
For convenience, these three solutions are
marked, respectively, with CFL slet-1, 2 and 3.
The ordinary strangelets without color-flavor locking
have also been plotted in the same figure for comparison purpose.
It can be seen that the charge of the CFL slet-1 is approximately
proportional to $A^{2/3}$, while that of the CFL slet-2 or 3 is
nearly proportional to $A^{1/3}$.

To have a better understanding of the three solutions, let's
take some mathematical analysis.

First, assume the common Fermi momentum $\nu$
is much bigger than the strange quark mass, i.e., $m_s/\nu\ll 1$.
In this case, one can take the limit of $m_{u,d}\rightarrow 0$
first, and then expand to a Taylor series with respect to $m_s$
on all the above expresions, to get simple expressions.
The expansion of the pressure is
\begin{eqnarray}
P&\approx&\frac{gm_s}{16\pi^2}\left[
     \frac{\pi\mu}{R^2}-\frac{4\nu(2\mu-\nu)}{R}
                   \right]
  +\frac{3\Delta^2}{\pi^2}\mu^2
          \nonumber\\
 &&  +\frac{g\nu}{8\pi^2}\left[
     \nu^2(4\mu-3\nu)-\frac{2\mu-\nu}{R^2}
                      \right]
  - B.
\end{eqnarray}
% Eq.\ (\ref{mubexp}) leads to
% \begin{equation}
%  \mu\approx
%   \nu+\frac{m_s^2}{6\nu}
%   -\frac{Rm_s^3}{3\nu(2R^2\nu^2-1)}.
% \end{equation}
For the quark number densities, they are
\begin{equation} \label{nud1}
n_{u,d}
\approx
 \frac{2\Delta^2}{\pi^2}\mu
 -\frac{g\nu}{4\pi^2R^2}
 +\frac{g\nu^3}{6\pi^2}
\end{equation}
and
\begin{equation}  \label{ns1}
n_s\approx
 \frac{2\Delta^2}{\pi^2}\mu
 -\frac{g\nu}{4\pi^2R^2}
 +\frac{g\nu^3}{6\pi^2}
 -\frac{3gm_s}{16\pi^2}
         \left[
  \frac{4\nu}{R}-\frac{\pi}{R^2}
         \right].
\end{equation}

It is obvious from Eqs.\ (\ref{nud1}) and (\ref{ns1}) that
$n_s$ is smaller than $n_{u,d}$ because $n_s$ has an extra
negative term. The corresponding strangelet,
CFL slet-1, is thus positively charged.

Secondly, assume $\nu$ is modest, i.e., it is smaller than $m_s$
but larger than $m_{u,d}$. In this case,
one can still take the limit of $m_{u,d}\rightarrow 0$ for $u/d$
quarks. But for $s$ quarks, expressions should be expanded according to
$\nu$, rather than $m_s$.
Accordingly, Eq.\ (\ref{eqP}) %% and (\ref{mubexp})
becomes  %% , respectively,
\begin{equation} \label{Papp2}
P=\frac{3\Delta^2}{\pi^2}\mu^2
  -\frac{gm_s\nu}{6\pi^2R^2}-B.
\end{equation}
%and
%\begin{equation}
%\mu\approx
%\nu+\frac{m_s-\nu}{2R^2\nu^2}.
%\end{equation}

The $u/d$ quark number density is still the same as Eq.\ (\ref{nud1}).
For $s$ quarks, however, one now has
\begin{equation} \label{ns2}
n_s\approx
 \frac{2\Delta^2}{\pi^2}\mu
 +\frac{3g\nu}{16\pi R^2}\left(
         \frac{8}{3}-\frac{\pi\nu}{ m_s}
                        \right)
 -\frac{3g\nu^2}{16\pi R}
\end{equation}

Please note the curvature (the $R^{-2}$ term) contribution.
It is negative for $u/d$ quarks in Eq.\ (\ref{nud1}).
However, it is positive for $s$ quarks in Eq.\ (\ref{ns2}).
This makes $n_s$ bigger than $n_{u,d}$.
Consequently, the corresponding strangelet, CFL slet-2,
is negatively charged.

Thirdly, assume $\nu$ is extremely small so that
expansion can be done with respect to $\nu$ for all the three flavors.
In this case, the pressure gives
\begin{equation} \label{Papp3}
P=\frac{3\Delta^2}{\pi^2}\mu^2
  +\frac{g}{2\pi^2}(\mu-\bar{m})\frac{\nu}{R^2}-B,
\end{equation}
where $\bar{m}\equiv (m_u+m_d+m_s)/3$,
while the quark number densities are
\begin{equation} \label{nappr}
n_u\approx n_d\approx n_s\approx
\frac{2\Delta^2}{\pi^2}\mu
 +\frac{g\nu}{2\pi^2R^2}.
\end{equation}
Namely, the three flavors of quarks are nearly equal in this case.
The corresponding strangelet, CFL slet-3, is almost neutral.

 \begin{figure}[ht]
\epsfxsize=8.2cm
\epsfbox{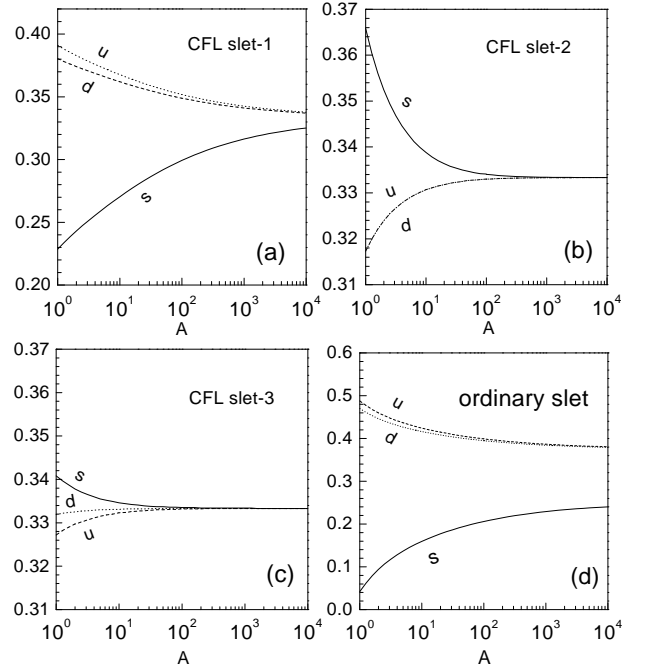}
 \caption{
Quark fractions of different strangelets.
Figures (a)-(c) are for the three kinds of CFL strangelets.
Figure (d) is for the ordinary strangelets.
The vertical axis for each figure is the quark number density
in unit of the total quark number density, or the ratio of the
corresponding quark number to the total quark number.
 \label{figna}}
 \end{figure}

Naturally, the above expanded expressions are merely approximate.
Real calculations have been performed by directly solving
the original system equations.
Fig.\ \ref{figna} shows the quark fraction
in different phases. They are qualitatively consistent with
the above analysis.
%A little difference from Fig.\ 2
%is that we take $m_u=5$ MeV and $m_d=10$ MeV, which are closer
%to the accepted current mass of light quarks \cite{Gasser1982PR87}.
%This makes $n_u$ and $n_d$ are not identical, and
%CFL slet-3 also gets some extremely small negative charge.

 \begin{figure}[htb]
\epsfxsize=8.2cm
\epsfbox{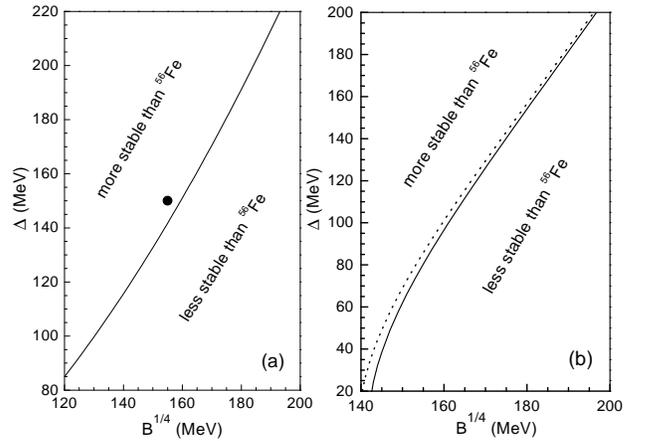}
 \caption{
Parameters for CFL strangelets to be more stable than $^{56}$Fe.
(a) is for CFL slet-2 and 3 while (b) is for CFL slet-1.
The full dot indicates the parameters in this paper.
 \label{figda}}
 \end{figure}

%Figure \ref{figea} is the energy density calculated by the
%normal thermodynamic relation $E=\Omega+\sum_i\mu_i n_i$.
%It indicates that the energy densities of CFL slet-2 and 3
%are very close to each other, and much lower than that of
%CFL slet-1 and the ordinary strangelets. Because quarks near the
%Fermi surface pair while the Fermi momenta of CFL slet-2 and 3
%are very small, nearly all quarks pair, especially in CFL slet-3.
% So the energy contributions are mainly from the pairing
%energy. Because two paired quarks looks like a boson,
%CFL slet-3 can be, to some extent, regarded as a 'Bose condensation'
%of Fermions.

Now we discuss the determination of parameters.
For the $u/d$ quark mass, we take $m_u=5$ MeV and $m_d=10$ MeV, which are
closer to the accepted current mass of light quarks \cite{Gasser1982PR87}.
Decreasing the $u/d$ quark mass has little effects
on CFL slet-1 and 2, while the charge of CFL slet-3
becomes smaller and smaller until it is charge-neutral.
The strange quark mass is expected to be density-dependent,
lying between the current mass $\sim$\ 100 MeV and the vacuum
constituent quark mass $\sim$\ 500 MeV.
%Because the concrete
%value are not as important as it is for the ordinary strangelets,
%we take $m_s=150$ MeV. % as in Ref.\ \cite{Madsen01PRL87}.
%The bag constant $B^{1/4}$ appeared in the
%Figs.\ 1 and 2 of Ref.\ \cite{Madsen01PRL87} is in the range of
%120 MeV to 200 MeV, and we take the modest value
%$B^{1/4}=155$ MeV
%which was also adopted in the Fig. 1 of Ref.\ \cite{Lugones2002PRD66}.
%A little later, we will see another interesting reason for taking
%this special value of the bag constant.
%As for the most important parameter is the gap $\Delta$.
The $\Delta$\ value varies from several tens to several
hundreds of MeV in literature. For example, it can range from 20
MeV to 90 MeV \cite{Rapp1998PRL81}, or from 50 MeV to more than
 100 MeV \cite{Alford1998PLB422}. Sophisticated
treatments of the instanton interaction, including form factors
from suitable Fourier transformation of instanton profiles,
give larger values for $\Delta$, as large as more than 200
MeV \cite{Carter1999PRD60}.
Therefore, we treat $\Delta$\ as a free parameter in the present
investigation.
For the above calculations in Figs.\ 2 and 3, we have taken
$\Delta=150$ MeV, $B^{1/4}=155$ MeV (This $B$ value was
also used in Ref.\ \cite{Lugones2002PRD66}), and $m_s=150$ MeV.
How these parameters influence the stability of CFL strangelets
will be discussed a little later.

Although the `common Fermi momentum' $\nu$\  in CFL slet-2 and 3 is
small, the chemical potential $\mu$ is still large.
%This is very important because, as mentioned in the beginning,
%CFL phase will give way to a new ``gapless CFL phase'' if the ratio
%$m_s^2/\mu$ is above a critical value about 47 MeV \cite{Alford04PRL92}.
%%From Eqs.\ (\ref{Pappr})-(\ref{nappr}), one can have
To get an approximate expression for $\mu$ from the equality $P=0$,
one can take $\nu=0$ in Eq.\ (\ref{Papp2}) or (\ref{Papp3}),
resulting
\begin{equation} \label{muappr}
\mu=\frac{\pi}{\sqrt{3}} \frac{\sqrt{B}}{\Delta}.
\end{equation}
For the parameters chosen for Figs.\ 2 and 3, Eq.\ (\ref{muappr}) gives
$\mu\approx 290$ MeV, very close to the actual value
from the numerical calculation, which gives the ratio $m_s^2/\mu$
to be about 77 MeV. % smaller than the critical value.
On the other hand, $\mu$\ varies in the range of $240-263$ MeV
for CFL slet-1. % The corresponding ratio is thus bigger,
%as shown in Fig.\ \ref{figms}.

The radius of CFL slet-2 and 3 can be approximately expressed as
\begin{equation} \label{R23}
R_{\mathrm{slet-2,3}}\approx
 \left(
 \frac{3\sqrt{3}A}{8\Delta\sqrt{B}}
 \right)^{1/3}.
\end{equation}
This equation means $R\propto A^{1/3}$,
which is a known fact in nuclear physics.
one may perhaps imagine $\Lambda$, composed of $(uds)$,
as the simplest CFL slet-3.
The H particle \cite{Jaffe76PRL38}, composed of $(uuddss)$,
is probably the next simplest CFL slet-3.
% In fact,
%the bag constant $B$ is chosen so that it gives a
%rough estimate of the $\Lambda$\ mass. If $H$ is really
%a CFL slet-3, its mass should be about
%two times that of $\Lambda$.

%If a strangelet is negatively charged and absolutely stable,
%and still negatively charged and absolutely stable
%after gorging with normal nuclear matter, catastrophe is imaginable:
%its runaway expansion will, in principle, consume our planet.
For information on the stability of CFL strangelets,
we should investigate the energy per baryon $E/n_{\mathrm{b}}$.
It is generally a function of $A, m_s, \Delta$,
 and $B$, i.e., $E/n_{\mathrm{b}}=f(A, m_s, \Delta, B)$.
If $E/n_{\mathrm{b}}$ is less than 930 MeV
(the mass of $^{56}\mbox{Fe}$ divided by 56),
the strangelets are absolutely more stable than normal nuclear matter.
Otherwise, they are meta-stable or unstable. The full line in
Fig.\ \ref{figda}(a) gives $\Delta$ as a function of
$B$ at $A=20$, $m_s=150$ MeV, $E/n_{\mathrm{b}}=930$ MeV.
In fact, this line does not
depend strongly on the concrete values of $m_s$ and $A$.
Because $E=\Omega+\sum_i\mu_i n_i=\Omega+3\mu n_{\mathrm{b}}$
and $\Omega\approx -P=0$, one has $E/n_{\mathrm{b}} \approx 3\mu$.
With a view to Eq.\ (\ref{muappr}), we immediately have
$ %\begin{equation}
 {E}/{n_{\mathrm{b}}}
\approx {\sqrt{3}\pi\sqrt{B}} / { \Delta }.
$ %\end{equation}
%Therefore, if $(B,\Delta)$ is located in the up-left part
%of Fig.\ \ref{figda}(a), the ratio $\sqrt{B}/Delta$\
%is less than $310\sqrt{3}/\pi\approx 170$ MeV,
%the new strangelets are absolutely stable.
Therefore, if
\begin{equation} \label{stabcond}
\frac{\sqrt{B}}{\Delta}
< \frac{310\sqrt{3}}{\pi}
\approx 170\ \mbox{MeV},
\end{equation}
then the parameter pair $(B,\Delta)$ is located in the up-left part
of Fig.\ \ref{figda}(a), and the new strangelets are more stable than iron.
For CFL slet-1, a similar solid line is plotted in Fig.\ \ref{figda}(b).
For different $m_s$ and $A$, this line moves a little up-left
(bigger $m_s$ e.g.\ the dotted line for $m_s=180$ MeV) or
down-right (smaller $m_s$).
However, the line in Fig.\ \ref{figda}(a)
 %  [dash line in Fig.\ \ref{figda}(b)]
is always located in the region where CFL slet-1 is
more stable than $^{56}$Fe for reasonable $m_s$. Therefore, if the condition
Eq.\ (\ref{stabcond}) is satisfied, all the three kinds of CFL
strangelets are more stable than $^{56}$Fe, and also
more stable than the normal unpaired strangelets.
As for the comparative stability between the three kinds of CFL
strangelets, it depends on the pairing parameter $\Delta$.
If one uses the same $\Delta$\ for all the three,
then the slet-1 is more stable. In this case, however,
the former is denser. Because investigations have shown that
$\Delta$ depends on dendity, most probably increases with
increasing densities \cite{Carter1999PRD60},
the comparative stability of the three kinds of CFL
strangelets needs to be further studied in the future.

CFL strangelets which are more stable than $^{56}$Fe may have far-reaching
consequencies. The slet-1 can provide an alternative explanation for cosmic
rays beyond the GZK cutoff \cite{Madsen03PRL90}.
The slet-3 is nearly neutral, and so might be a candidate for the
miracle dark matter in our universe.
%
%The slet-2 and 3 have lower
%ensities than that of slet-1, and closer to the normal nuclear
%saturation, and so may have more chances to be produced in the modern
%heavy ion collision experiments.
%However, it is unable to transform our planet into a strange star.
%When the electron's Compton wave length ($\approx 386$ fm)
%is reached, the constraint $n_e=0$, (or, equivalently,
%$\mu_u=\mu_d=\mu_s$) is no longer
%valid, and so the strangelet is neutralized and ceases to expand its size.
%Taking, for example, $\Delta=150$ MeV, $B^{1/4}=155$ MeV,
%and $R=386$ fm, Eq.\ (\ref{R23}) gives $A=4.2\times 10^7$.
%If on average the strangelet mass is, e.g., 58 MeV per baryon smaller
%than normal nuclei, the total energy released by the slet-2 would be
%about $58A\approx$\ 2409 TeV, much higher than the beam energy of
%the ALICE experiments at LHC being built at CERN.
%According to Ref.\ \cite{Dar}, the safe margin is proportional
%to ${E_{\mathrm{bm}}^{3.2}}$. Therefore, the risk for ALICE is
%probably a factor of $30^{3.2}\approx 53307$ higher than it is
%for RHIC whose beam energy is 30 times lower (20 TeV).
%
The slet-2 and 3 are more stable than the normal unpaired strangelets,
and so may have chances to be produced in the modern heavy ion collision
experiments. However, they are unable to transform our planet into a
strange star for the following two reasons. First, the positively charged
slet-1 is the energy minimum for the same parameters. And secondly,
when the electron's Compton wave length ($\approx 386$ fm) is reached,
the constraint $n_e=0$, (or, equivalently, $\mu_u=\mu_d=\mu_s$) is
no longer valid, and so the strangelet will be neutralized and ceases
to expand its size.

It should be emphasized that the strangelets reported here
are different from the previous ones \cite{Madsen00PRL85,Madsen01PRL87}
in that their electric charge is opposite. The new strangelets are
also different from the heavily negatively charged strangelets in
Ref.\ \cite{Schaf}. There the strangelets were not in
$\beta$\ equilibrium and had no color-flavor locking.
It was investigated how the metastable candidates
might look like if they are assumed to be stable against
strong hadronic decay and subsequently against weak hadronic decay.
Here the strangelets are assumed to be in perfect $\beta$\ equilibrium
and considered as having the possibility of absolute stability.
However, the concrete values should not be taken seriously, and
further studies are needed.

In summary, there exist new solutions to the system equations where
CFL strangelets are slightly negatively charged or nearly neutral.
If the ratio of the squared bag constant to the gap parameter is
smaller than 170 MeV, CFL strangelets are more stable than
the normal nuclear matter and ordinary unpaired strangelets.

The authors would like to thank support from
the DOE (DF-FC02-94ER40818),
NSFC (10375074, 90203004, and 19905011),
FONDECYT (3010059 and 1010976),
CAS (E-26), and SEM (B-122).
G.X.\ also acknowledges hospitality at MIT-CTP.
In particular, he is grateful to
K.\ Rajagopal and F.\ Wilczek for helpful discussions.

\end{document}